# SQUID-on-tip with single-electron spin sensitivity for high-field and ultra-low temperature nanomagnetic imaging

Y. Anahory[1*], H.R. Naren[2], E.O. Lachman[2], S. Buhbut Sinai[3], A. Uri[2], L. Embon[2], E. Yaakobi[2], Y. Myasoedov[2], M.E. Huber[4], R. Klajn[3] and E. Zeldov[2]

[1]Racah Institute of Physics, The Hebrew University, Jerusalem 91904, Israel

[2]Department of Condensed Matter Physics, Weizmann Institute of Science, Rehovot 7610001, Israel

[3]Department of Organic Chemistry, Weizmann Institute of Science, Rehovot 7610001, Israel

[4]Departments of Physics and Electrical Engineering, University of Colorado Denver, Denver 80217, USA

*Corresponding author: yonathan.anahory@mail.huji.ac.il

Scanning nanoscale superconducting quantum interference devices (nanoSQUIDs) are of growing interest for highly sensitive quantitative imaging of magnetic, spintronic, and transport properties of low-dimensional systems. Utilizing specifically designed grooved quartz capillaries pulled into a sharp pipette, we have fabricated the smallest SQUID-on-tip (SOT) devices with effective diameters down to 39 nm. Integration of a resistive shunt in close proximity to the pipette apex combined with self-aligned deposition of In and Sn, have resulted in SOT with a flux noise of 42 $n\Phi_0 Hz^{-1/2}$, yielding a record low spin noise of 0.29 $\mu_B Hz^{-1/2}$. In addition, the new SOTs function at sub-Kelvin temperatures and in high magnetic fields of over 2.5 T. Integrating the SOTs into a scanning probe microscope allowed us to image the stray field of a single $Fe_3O_4$ nanocube at 300 mK. Our results show that the easy magnetization axis direction undergoes a transition from the (111) direction at room temperature to an in-plane orientation, which could be attributed to the Verwey phase transition in $Fe_3O_4$.

Nanoscale magnetic imaging is gaining interest in recent years in diverse fields of research and applications, including biological, chemical, and physical systems[1–3]. The quest for quantitative mapping of extremely weak local magnetic signals has driven development of very sensitive new techniques, including cold atom chips[4], scanning nitrogen-vacancy diamond magnetometers[2,5–8], and scanning micro- and nano-superconducting quantum interference devices (SQUIDs)[9–15]. These tools have been employed for quantitative imaging of microscopic static and dynamic magnetic structures[6,16–20], current imaging[7,21,22], and responses to local perturbations[23,24]. Many of the systems of current interest in condensed matter physics, however, including topological states of matter, fractional quantum Hall, spin ice, and unconventional superconductivity, often require imaging at high spatial resolution with high spin sensitivity at sub-Kelvin temperatures and at high magnetic fields. These conditions are currently inaccessible by most of the existing imaging techniques. In this work, we present the development of scanning nanoSQUIDs that operate at sub-Kelvin temperatures and significantly expand the state-of-the-art at these temperatures in terms of spatial resolution, spin sensitivity, and applied magnetic fields.

Previously, we have developed a highly sensitive scanning probe microscope based on a nanoscale SQUID that resides at the apex of a sharp tip (SQUID-on-tip, or SOT)[14,25,26] and utilized the technique for the study of vortex dynamics[19,27], nanoscale magnetism in topological insulators[28,29], and superparamagnetism at oxide interfaces[18]. The SOT devices made of Pb attain loop diameters down to 46 nm and display extremely low flux noise of 50 $n\Phi_0 Hz^{-1/2}$ and record spin sensitivity of 0.38 $\mu_B Hz^{-1/2}$ at 4.2 K[14]. In addition, the Pb SOTs show an extremely high thermal sensitivity of below 1 $\mu K \cdot Hz^{-1/2}$, allowing nanoscale imaging of dissipation mechanisms in quantum systems. Such imaging reveals the minute quantity of heat generated by inelastic scattering of electrons from a single atomic defect[30,31]. At sub-Kelvin temperatures, however, the Pb SOTs become highly hysteretic due to an increase in the critical current $I_c$ and thermal instabilities. As a result, these SOTs with outstanding sensitivity at 4.2 K are barely useable at ultra-low temperatures. SOTs made of Al were shown to operate at sub-Kelvin temperatures[25], but they display a significantly higher noise of 2 $\mu\Phi_0 Hz^{-1/2}$ due to high kinetic inductance and their operation is limited to magnetic fields of below ~0.5 T.

To improve the sensitivity, spatial resolution, and operating range of temperatures and magnetic fields of the SOT, we describe here a number of technological advancements, including *i*) incorporation of grooved quartz pipettes, *ii*) reduction of SOT diameter, *iii*) integration of a resistive shunt on the tip and *iv*) development of SOTs made of In and Sn.

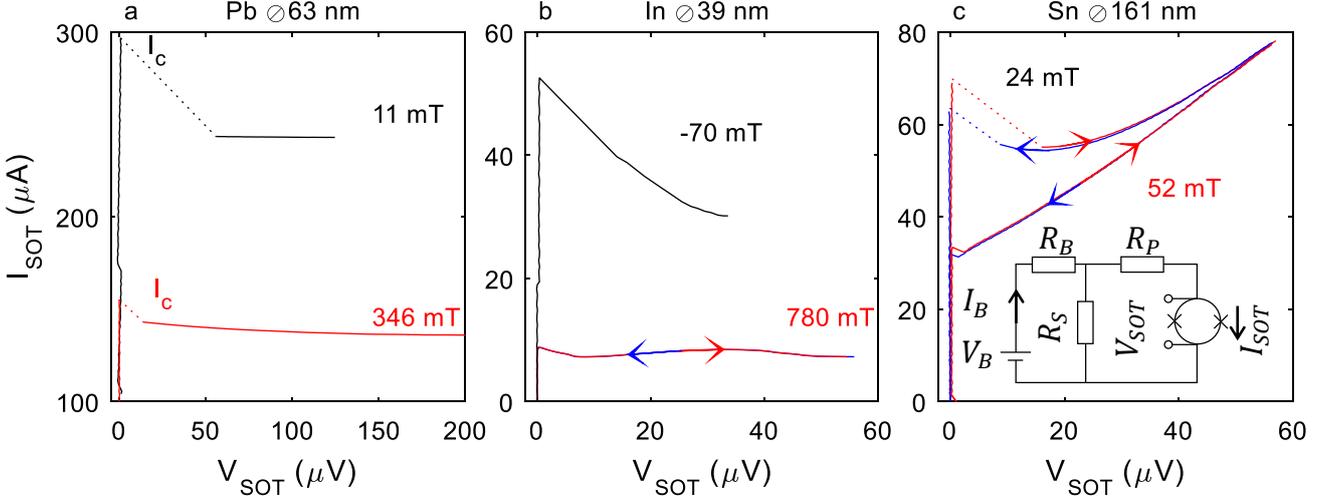

**Figure 1.** *I-V* **characteristics of SOT devices made of different materials at 300 mK.** (a-f) *I-V* characteristics of three different SOT devices at two different applied magnetic fields. (**a**) $I_{SOT}$-$V_{SOT}$ curve of a hysteretic unshunted Pb SOT with diameter of 63 nm ($R_B$ = 2.0 kΩ, $R_s$ = 3.0 Ω, $R_P$ = 3.1 Ω). (**b**) $I_{SOT}$-$V_{SOT}$ characteristics of an In SOT of 39 nm diameter with integrated shunt-on-tip ($R_B$ = 7.0 kΩ, $R_s$ = 3.0 Ω, $R_P$ = 1.8 Ω). The red (blue) curve shows the measurement upon ramping $I_B$ up (down). (**c**) $I_{SOT}$-$V_{SOT}$ characteristics of a slightly hysteretic shunted Sn SOT with diameter of 161 nm ($R_B$ = 7.0 kΩ, $R_s$ = 3.0 Ω, $R_P$ = 1.8 Ω). The inset shows a simplified diagram of the measurement circuit.

To illustrate the limitations of Pb SOTs at ultra-low temperature, we inspect their current-voltage (*I-V*) characteristics. Figure 1a shows typical $I_{SOT}$-$V_{SOT}$ curves of a Pb SOT at two different applied fields at 300 mK. The measurement circuit, illustrated schematically in the inset of Fig. 1c, consists of a room temperature voltage source that provides the bias current $I_B$ through the low temperature bias resistor $R_B$ with typical resistance of a few kΩ. The SOT is effectively voltage biased by the low resistance shunt resistor $R_s$ of a few Ω in presence of a parasitic resistance $R_p$. More data and details about the measurements are shown in the electronic supplementary information (ESI). The $I_{SOT}$-$V_{SOT}$ curves can be separated into two parts. Below $I_c$, the SOT is superconducting and $V_{SOT}$ is zero, while for larger values of $I_{SOT}$, the SOT is in the voltage state. The $I_{SOT}$-$V_{SOT}$ curve of the SOT in Fig. 1b shows that, once the critical current $I_c$ is reached, the voltage jumps discontinuously to a large value of 55 μV (dashed line) with $I_{SOT}$ showing almost no dependence on the bias above $I_c$. In this hysteretic state, the differential resistance ($dV_{SOT}/dI_{SOT}$) is too high to be measured accurately with our circuit. Its value, on the order of several tens of ohms, indicates that the SQUID loop and part of the leads are in the normal state. In this state, the SOT cannot serve as a sensitive magnetometer. To circumvent this problem, the SOT characteristics must be damped to attain non-hysteretic behavior.

An important parameter to consider for identifying the cause of the hysteresis is the Stewart-McCumber parameter $\beta_c = 2\pi I_c R^2 C / \Phi_0$, where $R$ and $C$ are the normal resistance and capacitance of the Josephson junction, respectively, and $\Phi_0$ is the flux quantum. This parameter is derived from the resistively and capacitively shunted junction (RCSJ) model[32] and provides a qualitative description of the SOT. For $\beta_c > 1$, the Josephson junctions are underdamped and likely to be hysteretic. However, when $β_c < 1$, the desired overdamped non-hysteretic characteristics are expected. For $I_c = 300$ μA and an estimated $C \cong 2$ fF and $R \cong 50$ Ω, we obtain $β_c \cong 5$ for the Pb SOT in Figs. 1a,b. This result indicates that the $β_c$ parameter should be reduced by a factor of ~10 by decreasing $R$ and/or $I_c$ to prevent hysteresis.

We first consider the possibilities of reducing $I_c$. $I_c$ can be reduced *in-situ* by increasing the temperature. Indeed, we find that the Pb SOTs hysteretic at 300 mK usually become non-hysteretic above 4 K. Here, however, our goal is to attain sensitive operation at ultra-low temperatures. The second possibility for *in-situ* reduction of $I_c$ is the application of high magnetic fields. As we show below, the Pb SOT indeed becomes non-hysteretic at fields above ~1.5 T. This approach, however, is not applicable if low- or variable-field operation is required.

The most straightforward means for reducing $I_c$ is reducing the thickness of the superconducting film forming the SQUID loop. This approach, however, does not work reliably for Pb films, which show significant surface roughness, and therefore requires a minimum film thickness to attain superconducting percolation. We therefore aim to employ other superconducting materials with a lower critical temperature $T_c$ in order to attain lower $I_c$ at sub-Kelvin temperatures. We have accordingly developed a technique to deposit In and Sn thin films onto the quartz pipette. In order to overcome the high room-temperature surface mobility of In and Sn, which results in island growth rather than a continuous thin film, we utilized a high-vacuum system with an *in-situ* rotatable $^4$He cryostat for thermal deposition of In and Sn onto cryogenically cooled tips. This technique is similar to that used to make Pb SOTs and is discussed in Ref. 14. Scanning electron micrographs of the resulting In and Sn SOTs are shown in Figs. 2e-g.

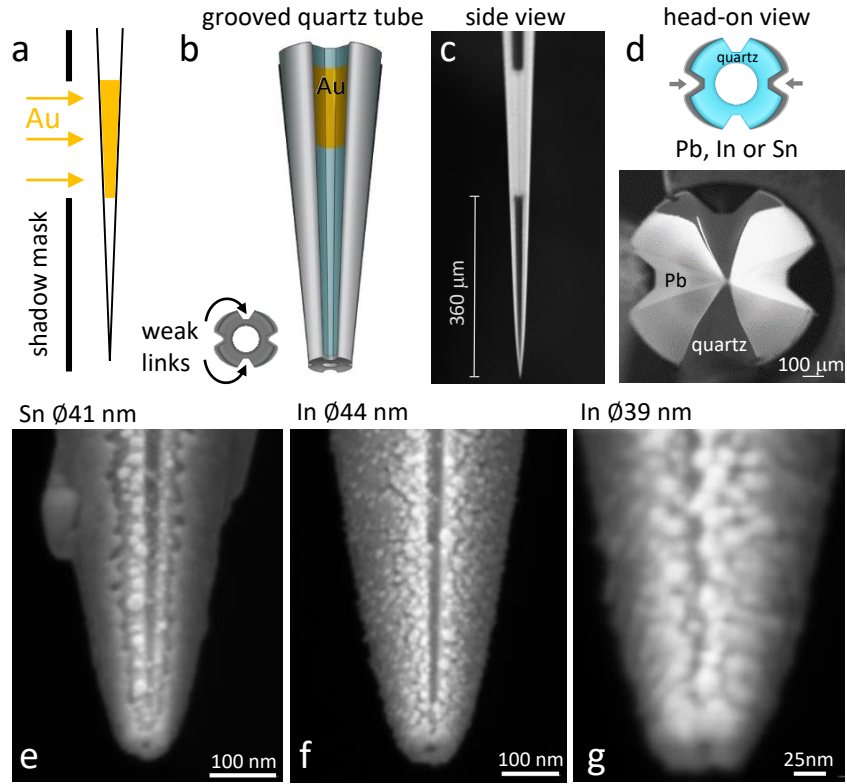

**Figure 2. Fabrication of shunted Sn and In SOT devices on grooved pipettes.** (**a**) Schematic drawing of the fabrication process of the shunt-on-tip using Au deposition through an aperture in a shadow mask. (**b**) Schematic drawing of the final SOT structure. The pipette is fabricated by laser pulling of a quartz capillary with four grooves along its length. Following deposition of the shunt-on-tip, two superconducting leads are deposited on two opposite sides of the pipette (grey) and a third head-on deposition is used to coat the apex of the pipette, forming the SQUID loop. The inset shows the apex viewed from below coated with a superconductor (grey) with two weak links formed at the constrictions. (**c**) SEM side view image of the final SOT structure showing the metallic layers in bright (Au shunt-on-tip and the In side electrodes) and the bare quartz in dark. The length of the shunt-on-tip is 250 μm and it is located 360 μm away from the tip apex (shown in (e-g)). (**d**) Schematic cross section of the quartz capillary (blue) with superconducting leads (grey) on the two sides (top). Low-magnification SEM head-on image (bottom) showing the grooved pipette (1 mm outer diameter) with Pb leads on two opposite sides (bright) extending all the way to the apex in the center. (**e-g**) SEM images of different SOT devices made of Sn (**e**) and In (**f,g**) with effective diameters of 41 (e), 44 (f), and 39 nm (g). The quantum interference patterns of devices (f) and (g) are shown in Figs. 3b and 3c respectively.

An additional approach for reducing $\beta_c$ is the reduction of $R$. In fact, this method is expected to have a larger influence on $\beta_c$ since $\beta_c \propto R^2$. Since the normal state resistance of the superconducting film cannot be varied independently of $I_c$, we incorporate a shunt resistor integrated with the SOT. At the relevant Josephson frequencies ($\gtrsim 1$ GHz), an inherent impedance mismatch exists between the SOT and the long ($\sim 1$ cm) and narrow leads along the pipette; any external shunt resistor would have little effect due to this high-series impedance. To avoid this situation, the shunt resistor must be incorporated on the pipette in close vicinity to the SQUID loop. Depositing a normal metal directly onto the apex of the tip could reduce $R$. However, because of the small loop size, it is not possible, on the apex itself, to achieve a thickness that would reduce the normal resistance significantly. Moreover, in this configuration, we found that the inverse proximity effect[33] significantly suppresses the superconductivity in the SQUID loop. Taking into account these design limitations, we have developed a shunt-on-tip as shown in Fig. 2. Using a shadow mask (Fig. 2a), a thin film of Cr/Au (~5 nm/~10 nm) is deposited on the bare quartz pipette forming a strip ~250 μm long at a distance of ~400 μm from the apex (Fig. 2b,c). After the subsequent self-aligned deposition of the superconducting leads (Fig. 2d), this metallic film acts as a resistive shunt between the two leads. By tuning the thickness and the length of the Cr/Au film the desired resistance in the range of 2 to 10 Ω can be attained. We find that although the sole addition of the shunt-on-tip could not solve the hysteresis problem for a Pb SOT ($I_c \sim 250$ μA), it is a crucial design and fabrication parameter to control undesirable hysteresis specifically with Sn and In SOTs and with Pb SOTs at elevated fields when $I_c \lesssim 100$ μA. The shunt-on-tip also has the added benefit of protecting the SOT from static discharge while the device is at room temperature.

Sn and In SOTs fabricated with the integrated shunt-on-tip as described above were characterized at 300 mK. Typical $I_{SOT}$-$V_{SOT}$ curves are shown in Figs. 1a-c. Their critical currents are about five times lower than in Pb devices (Fig. 1a). In addition, the Au shunt-on-tip lowers significantly the normal state resistance. For example, the Sn SOT in Fig. 1c shows a differential resistance of 0.9 Ω at high bias. The combination of a lower $I_c$ and reduced $R$ eliminates the hysteresis as shown by the ramp up (red) and ramp down (blue) characteristics in Figs. 1c.

An additional goal of this work was to reduce the SOT diameter, which has two advantages. The first is improved spatial resolution for scanning magnetic imaging. The second is reduced spin noise, aiming at attaining single-electron spin sensitivity or better. A SQUID is fundamentally characterized by its flux noise $S_\Phi^{1/2}$. The corresponding field noise in the case of a uniform magnetic field is given by $S_B^{1/2} = S_\Phi^{1/2}/(\pi r^2)$, which decreases as $1/r^2$, where $r$ is the SQUID radius. Large SQUIDs are therefore preferable for sensitive measurements of uniform magnetic fields. Scanning nanoscale microscopy, however, requires imaging of spatially varying magnetic fields and resolving their local sources, in particular magnetic dipoles. In this case, the relevant performance metric is magnetic dipole, or spin, sensitivity. Assuming that the spin is located in the center of the SQUID, the spin sensitivity in units of $\mu_B \cdot Hz^{-1/2}$ is given by $S_n^{1/2} = S_\Phi^{1/2} r/r_e$, where $S_\Phi^{1/2}$ is in units of $\Phi_0/Hz^{1/2}$ and $r_e = 2.82 \times 10^{-15}$ m is the classical electron radius[34]. Since $S_n^{1/2} \propto r$, smaller SQUIDs achieve a lower spin noise.

The limiting factor in the fabrication of ultra-small SOTs is maintaining a gap between the superconducting leads along the pipette. SOT fabrication is based on a self-aligned process in which the superconducting leads are formed by two side-depositions, followed by a third head-on deposition step coating the tip apex[14,25]. In conventional uniformly circular capillaries, as the diameter of the pipette is reduced, the gap size is reduced proportionally, leading eventually to formation of shorts and additional Josephson junctions between the leads. In order to maintain a sufficient gap between the leads, we have employed a quartz capillary with four grooves[35] along its length as depicted in Fig. 2b,d. These grooves maintain their original shape during the pipette pulling process and provide shadowing during the first two deposition steps, resulting in a well-defined gap down to smaller pipette diameters than is possible for a uniform capillary. For this purpose, two grooves on the opposite sides of the pipette would have been sufficient. However, pulling a capillary with two grooves results in a pipette with an oval cross section. Therefore, in order to maintain a circular shape, four symmetric grooves were used as shown in Fig. 2. Employing this approach, we have attained a nanoSQUID with a record small effective diameter of 39 nm depicted in Fig. 2g.

The grooves have the additional advantage of reducing the effective width and length of the two weak link nano-constrictions (Fig. 2b inset). Tuning the nano-constrictions dimensions could be used to tune $I_c$ and $\beta_c$, as well as the kinetic inductance of the junctions for better modulation depth and improved flux noise. Thus, the depth of the grooves and their shape can be designed to attain optimal performance of the SOT.

The quantum interference patterns of the various SOT devices, measured at 300 mK, are presented in Fig. 3. The Sn SOT with an effective diameter of 161 nm shows a number of oscillation periods with an envelope of decreasing $I_c$ that vanishes at the upper critical field $H_{c2}$ of about 0.4 T (Fig. 3a). The period of the quantum oscillations increases with decreasing SOT diameter. As a result, in the smaller In devices, only the central lobe of the interference pattern is fully visible (Figs. 3b,c), while the following lobes are truncated by $H_{c2}$. The In SOT presented in Fig. 2g shows a quantum oscillation period of $\mu_0 \Delta H_z = 1.69$ T (Fig. 3c). From this period we calculate the effective diameter $d = \left(4\Phi_0/\pi\mu_0\Delta H_z\right)^{1/2} = 39$ nm. This constitutes the smallest SOT reported to date[14].

The operation of many SQUIDs is limited to relatively low fields due to penetration of vortices into the SQUID loop. In SOTs, in contrast, the width of the SQUID loop is usually smaller than the coherence length and the leads are aligned essentially parallel to the applied field, preventing vortex penetration. As a result, the SOTs can operate in applied fields of up to $H_{c2}$. The 44 nm diameter In SOT in Fig. 3b shows an interference pattern up to a record high field of 1.9 T. Even higher field operation can be attained with Pb SOTs. At 300 mK, the Pb SOTs are not useable below about 1.5 T due to too high critical current causing hysteretic behavior. However, as a result of the suppression of $I_c$ with magnetic field, operation in fields of up to almost 3 T can be attained as demonstrated in Fig. 3d for $d = 65$ nm Pb SOT. Here we show for clarity the tip response to magnetic field, $dI_{SOT}/d(\mu_0 H_z)$, which is directly related to the device sensitivity.

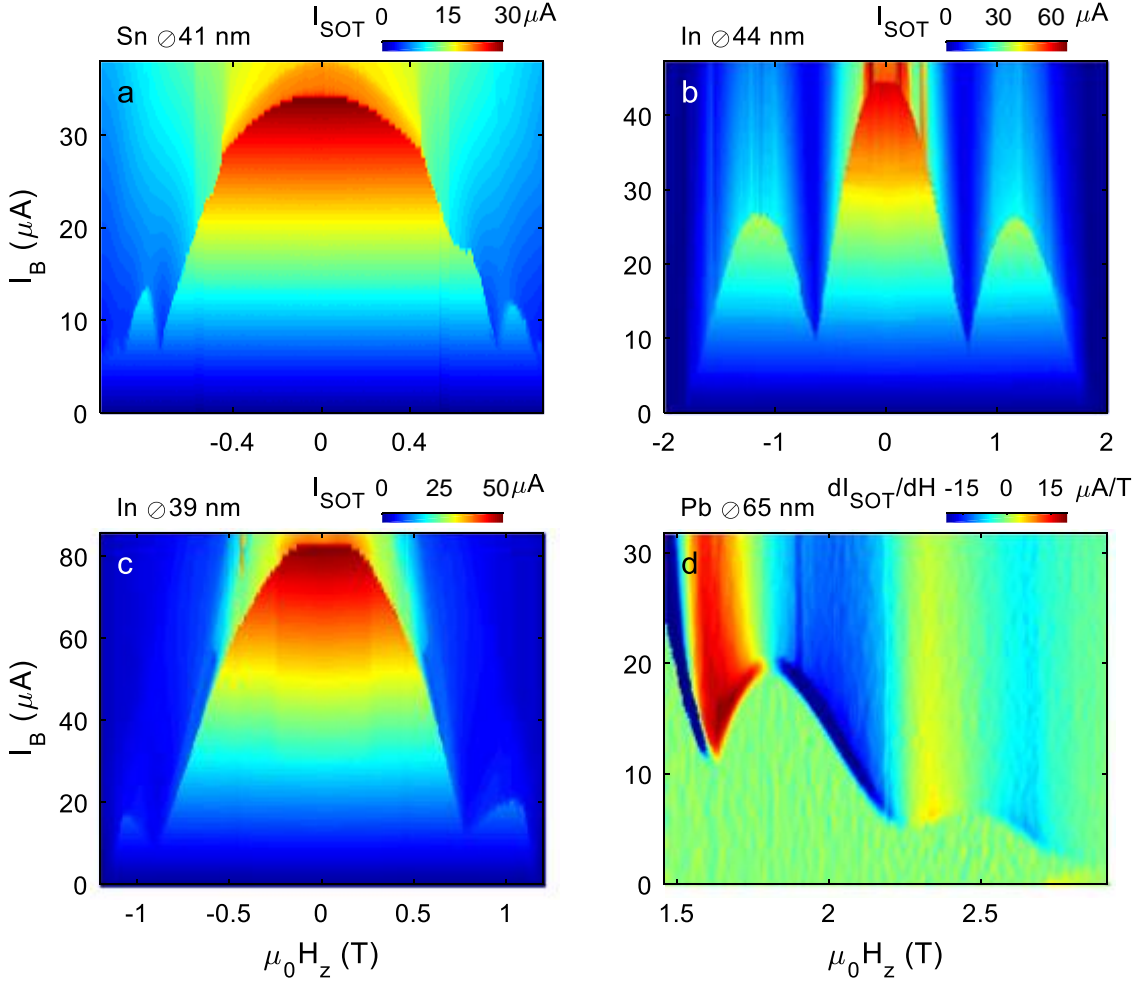

**Figure 3. Quantum interference patterns of four SOT devices at 300 mK.** (**a-c**) Color rendering of $I_{SOT}$-$V_B$ characteristics vs. applied field $\mu_0 H_z$. (**a**) Sn SOT with an effective diameter $d = 41$ nm ($\mu_0 \Delta H_z = 1.53$ T) measured using the circuit in Fig. 1f with $R_B = 7.0$ kΩ, and $R_S = 3.0$ Ω. (**b**) Same as (a) for an In SOT with $d = 44$ nm ($\mu_0 \Delta H_z = 1.37$ T), $R_B = 6$ kΩ, and $R_S = 1$ Ω. (**c**) Same as (a) for an In SOT of $d = 39$ nm ($\mu_0 \Delta H_z = 1.69$ T), $R_B = 7.0$ kΩ, and $R_S = 3.0$ Ω. (**d**) Derivative of $I_{SOT}$ with respect to $\mu_0 H_z$ presenting the response function of a Pb SOT with $d = 65$ nm ($\mu_0 \Delta H_z = 0.63$ T) at elevated fields, $R_B = 11$ kΩ, and $R_S = 1.4$ Ω).

The flux sensitivity of the SOT is determined by dividing the current spectral noise density in units of AHz$^{-1/2}$ by the transfer function $dI_{SOT}/d\Phi$, where $\Phi$ is the magnetic flux penetrating through the SQUID loop. The resulting flux noise is shown in Fig. 4 for the 39 nm device at two different fields, 0.74 T (red) and 1.1 T (blue). Above 3 kHz, the noise becomes essentially frequency-independent and we attain flux noise as low as $S_\Phi^{1/2} = 42$ n$\Phi_0$Hz$^{-1/2}$ at 0.74 T and 45 n$\Phi_0$Hz$^{-1/2}$ at 1.1 T. Remarkably, this device reaches a spin sensitivity of 0.29 and 0.31 $\mu_B$Hz$^{-1/2}$ at 0.74 and 1.1 T respectively, which makes it the most sensitive SQUID in terms of spin noise reported to date[14].

The In device with $d = 44$ nm shown in Figs. 2c and 3b also achieved impressive flux and spin sensitivity values of 145 n$\Phi_0$/Hz$^{1/2}$ and 1.1 $\mu_B$Hz$^{-1/2}$ respectively at a field of 1.87 T. The Sn SOTs did not show any significant advantage over In SOTs in terms of sensitivity or operating magnetic fields. Nonetheless, the use of Sn may prove to have practical advantage since it oxidizes less than In and hence Sn SOTs may have a longer shelf life.

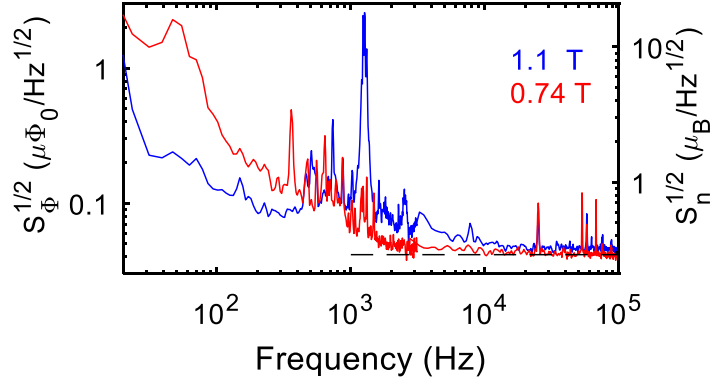

**Figure 4. Flux and spin noise spectral densities of ⌀39 nm In SOT at 300 mK.** The $S_\Phi^{1/2}$ (left axis) and $S_n^{1/2}$ (right axis) are shown for applied field of 0.74 T (red) and 1.1 T (blue). The excess noise around 1 kHz is due to experimental setup. In the white noise region above 3 kHz the flux noise reaches 42 $n\Phi_0 Hz^{-1/2}$ and the spin noise 0.29 $\mu_B Hz^{-1/2}$, which is the lowest ever reported for a SQUID.

To demonstrate the utility of the improved devices, an In SOT was used to investigate the low-temperature magnetic properties of single magnetite ($Fe_3O_4$) nanocubes. Bulk measurements at room temperature have shown that the easy magnetization axis of magnetite is in the $\langle 111 \rangle$ direction[36,37], which corresponds to the body diagonals of the cube[38,39]. Below $T_V \sim 120$ K, there is an abrupt change in many of the physical properties of the material such as the electrical conductivity, specific heat, lattice structure, and magnetization[40]. This so-called Verwey transition is believed to be caused by a long-range charge order[41,42] and has motivated many studies of size effects in nanocrystals[43–46] and thin films[47,48] below $T_V$. Apart from an interest in the Verwey transition, there is a fundamental and technological interest in probing the effects of shape and size on magnetism[17,49]. However, most magnetic characterization studies have been limited to large particle ensembles[50,51] or to non-spatially resolved magnetic signals[52–57].

In order to probe the properties of individual nanocubes, a 150 nm diameter In SOT was integrated into a home-built scanning probe microscope operating at 300 mK[58] with the aim of resolving the spatial distribution of the out-of-plane component of the stray magnetic field $B_z(x,y)$. SQUIDs usually have regions of reduced or zero sensitivity ("blind spots") at the interference pattern extrema. As a result, a SQUID with two perfectly symmetric arms is not sensitive at zero field. In this experiment, a larger device was used in order to yield a smaller interference pattern period, which translates into denser sensitive regions in the field domain. Combining that with an inherent device asymmetry leads to good sensitivity at zero applied field. Smaller SOTs (down to 43 nm) were used in other microscopy experiments recently published[59,60] using the same improvements discussed in this work. Nanocubes of nominal average size $l \cong 13$ nm were synthesized as described in Ref. 39 and dropcasted onto a Si substrate. Figures 5a-d show the $B_z(x,y)$ images acquired at four different applied fields $\mu_0 H_z$. Prior to the first measurement, the magnetic field was raised to $\mu_0 H_z = 0.36$ T to calibrate the SOT and then decreased to zero. At $\mu_0 H_z = 0$ (Fig. 5a), the magnetization of the cube is almost perfectly aligned with the substrate plane. If the nanocube was magnetized along its diagonal axis, as at room temperature, this would imply that the cube is standing on one of its vertices, which is highly unlikely considering the strong van der Waals forces between the cube face and the substrate[61] – indeed, high-resolution SEM imaging (Fig. 5l) confirmed that nanocubes are found with one face laying on the substrate[39]. We therefore conclude that at $\mu_0 H_z = 0$ the magnetization of the nanocube at 300 mK occurs in the {100} plane (cube face) and not in the $\langle 111 \rangle$ direction. Indeed, for bulk magnetite, it is known that for $T < T_V$, the easy magnetization axis changes from $\langle 111 \rangle$ to $\langle 100 \rangle$[37,40]. It was also shown that nanocubes synthesized by the same method are magnetized in the <111> direction at room temperature[38]. Our nanoscale magnetic imaging thus provides a direct observation of the axis of preferential magnetization of a single, isolated nanocube below the Verwey transition.

Some nanocubes could have defects that may influence their magnetic state. We imaged four other nanocubes in this study (see ESI, Section 2) and found an in-plane magnetization of comparable magnitude. We did not observe any other magnetization orientation at zero-field, which is somewhat surprising since a purely out-of-plane magnetization would also be consistent with magnetization in the <100> direction. We speculate that an interaction with the underlying substrate is responsible for breaking this symmetry.

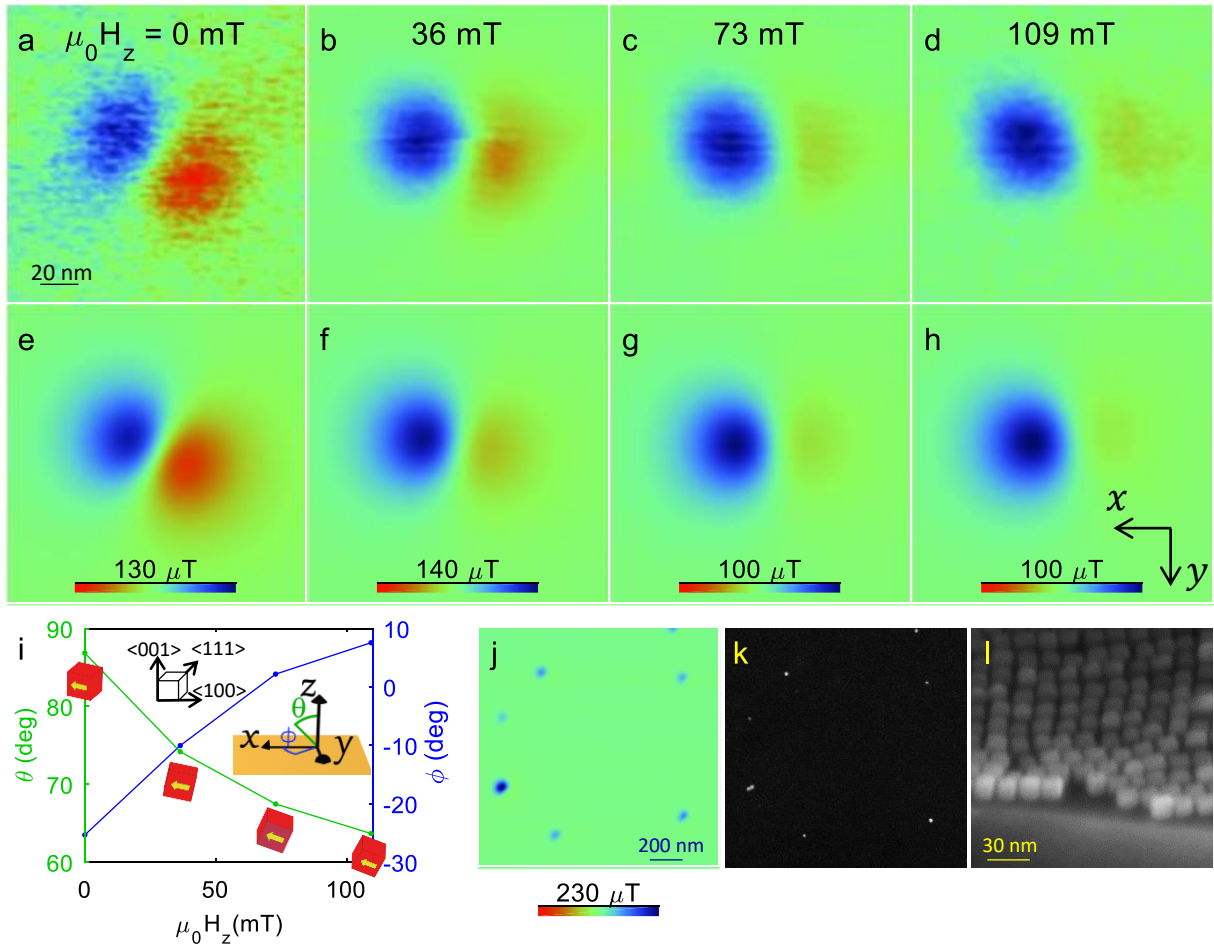

**Figure 5. Low-temperature characterization of the magnetization of a single $Fe_3O_4$ nanocube.** (**a-d**) Stray field $B_z(x,y)$ of a single $Fe_3O_4$ 13 nm nanocube at $\mu_0 H_z = 0$ T (**a**), 36 mT (**b**), 73 mT (**c**), and 109 mT (**d**) at 300 mK. (**e-h**) Corresponding best fit numerically simulated $B_z(x,y)$ with a constant magnetic moment $m_0 = 2.3 \times 10^5$ $\mu_B$ and varying azimuthal and polar angles. (**i**) Polar angle $\theta$ (green) and azimuthal angle $\phi$ (blue) and as a function of the applied field as derived from the fitting procedure. The cubes in red indicate the out-of-plane tilting and rotation of the nanocube upon increasing the applied field while keeping the magnetization orientation fixed to the cube axes (yellow arrow). Right inset: coordinate system with respect to the substrate (orange). Top inset: the crystallographic directions of a nanocube. (**j**) Measured $B_z(x,y)$ of a few nanocubes on a Si substrate at 4.2 K and $\mu_0 H_z = 0.4$ T. The darker blue spot on the left corresponds to two nanocubes next to each other. (**k**) SEM image of the same field of view as in (j) showing the nanocubes in bright. (**l**) High-resolution SEM image of an array of nanocubes viewed at an angle.

Figures 5b-d show the measured $B_z(x,y)$ at $\mu_0 H_z = 36$, 73, and 109 mT, respectively. For each of the images, the magnitude of the magnetic moment $m_0$ as well as its polar, $\theta$, and azimuthal, $\phi$, angles were derived by performing a numerical fit of the data. Our model comprises a finite size SQUID loop with fitting parameters $m_0$, $\theta$, $\phi$, and the SOT-to-nanocube distance. The corresponding numerically attained best-fit stray field maps are presented in Figs. 5e-h. We derive a magnetic moment of $m_0 = 2.3 \times 10^5$ $\mu_B$, which is quite consistent with the volume magnetization reported for $Fe_3O_4$ nanoparticles at low temperature[50]. A larger-scale $B_z(x,y)$ image of seven nanocubes is shown in Fig. 5j. According to the SEM image of the same area (Fig. 5k), five nanocubes are well separated and two nanocubes on the left are bunched together, giving rise to the higher $B_z(x,y)$ signal (darker blue in Fig. 5j).

The derived angles $\theta$ and $\phi$ are shown in Fig. 5i. The out-of-plane component of the magnetization increases with increasing $H_z$, which can be attributed to either rotation of the magnetization orientation within the nanocube, or the rotation of the entire nanocube. The work $W$ performed on the nanocube by the magnetic field from $\mu_0 H_z = 0$ to 0.109 T is $W = \mu_0 \int_{\theta_1}^{\theta_2} \vec{m} \times \vec{H} d\varphi = 3 \times 10^{-20}$ J or 6 kJ/m³, which is smaller than the magnetocrystalline anisotropy required for the rotation of magnetization orientation[62]. This fact, combined with the continuous variation of the angles upon ramping the field up and down, suggests that, despite the van der Waals force that acts as a restoring force, the entire nanocube is tilted out of plane upon increasing the field rather than that the magnetization orientation is varied within a stationary nanocube. The azimuthal angle $\phi$ is also observed to change slightly with $H_z$. Interestingly, upon reducing $H_z$ back to zero, the magnetic moment returned to the in-plane orientation, $\theta \sim 90°$, but with a slightly different azimuthal angle, indicating that the nanocube "landed" with a slightly different in-plane orientation after the out-of-plane field-induced tilting. If the nanocube stayed still, we would expect the azimuthal angle of the magnetization to change by a multiples of 90°.

In conclusion, we have developed SQUID-on-tip devices made of In and Sn and have introduced grooved pipette geometry and an integrated resistive shunt on tip. These advances led to the smallest size In SOTs of 39 nm effective diameter, ultra-low flux noise of 42 n$\Phi_0$Hz$^{-1/2}$, record spin sensitivity of 0.29 $\mu_B$Hz$^{-1/2}$, operation at ultra-low temperatures, as well as enhanced operating field of Pb SOTs up to over 2.5 T. The new devices were utilized to image the stray magnetic field generated by a single superparamagnetic $Fe_3O_4$ nanocube, revealing that the easy magnetization axis resides in the {100} plane rather than in the <111> direction and providing a direct manifestation of the Verwey phase transition observed on a single nanocube level. The new generation of the SQUID-on-tip devices paves the way to nanoscale magnetic and thermal imaging of a wide range of quantum and magnetic systems.

**Acknowledgements**

The authors thank M. L. Rappaport for technical assistance. This work was supported by the European Research Council (ERC) under the European Union's Horizon 2020 research and innovation program (grants No 785971 and 802952), by the US-Israel Binational Science Foundation (BSF) (grant No 2014155), and by the Weston Nanophysics Challenge Fund. EZ acknowledges the support of the Leona M. and Harry B. Helmsley Charitable Trust grant 2018PG-ISL006.

# Supplementary information

## SQUID-on-tip with single-electron spin sensitivity for high-field and ultra-low temperature nanomagnetic imaging

Y. Anahory[1*], H.R. Naren[2], E.O. Lachman[2], S. Buhbut Sinai[3], A. Uri[2],
L. Embon[2], E. Yaakobi[2], Y. Myasoedov[2], M.E. Huber[4], R. Klajn[3] and E. Zeldov[2]

[1]Racah Institute of Physics, The Hebrew University, Jerusalem 91904, Israel

[2]Department of Condensed Matter Physics, Weizmann Institute of Science, Rehovot 7610001, Israel

[3]Department of Organic Chemistry, Weizmann Institute of Science, Rehovot 7610001, Israel

[4]Departments of Physics and Electrical Engineering, University of Colorado Denver, Denver 80217, USA

*Corresponding author: yonathan.anahory@mail.huji.ac.il

## Measurement of I-V characteristics

The measurement circuit, illustrated schematically in Fig. S1, consists of a room temperature voltage source that provides the bias current $I_B$ through the low temperature bias resistor $R_B$, with a typical value of a few kΩ. The SOT is effectively voltage-biased by the low-resistance shunt resistor $R_s$, with a value of a few Ω, in presence of a parasitic resistance $R_p$. The current through the SOT, $I_{SOT}$, was measured using a SQUID series array amplifier (SSAA) as described in Refs.[1,2] and indentified in Fig. S1 as A. The SSAA is inductively coupled to the SOT circuit through a superconducting coil $L_{in}$. Warm electronics applies a voltage $V_{FB}$ to maintain a constant flux in the SSAA. In flux-locked loop conditions, $I_{FB} \propto I_{SOT}$ and $I_{FB}$ is itself proportional to the voltage drop $V_{FB}$ on a resistor $R_{FB}$, with a typical value of a few kΩ. Thus, $V_{FB}$ is a direct readout of $I_{SOT}$.

Figure S2a,c,e shows the raw $I_{SOT}$-$I_B$ curves of a SOTs at 300 mK for different materials. The $I_{SOT}$-$I_B$ curves can be separated into two regions. Below $I_c$, a linear region with slope $R_s/(R_s + R_p)$ represents the state where the SOT is superconducting. For larger values of $I_B$, the SOT is in the voltage state and the curve is no longer linear. Knowing the voltage on the shunt resistor $V_S = R_S(I_B - I_{SOT})$, and the voltage on the parasitic resistance $V_p = R_p I_{SOT}$, we calculate $V_{SOT} = V_S - V_P$. From this result, we can then present the $I_{SOT}$-$V_{SOT}$ curves (Fig. S2 b,e,f).

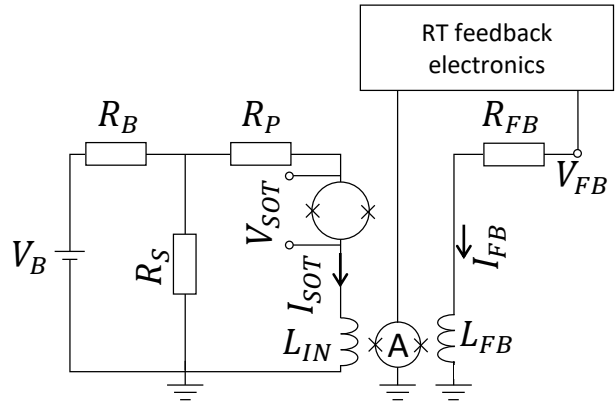

**Figure S1 – Detailed measurement circuit.** $R_B$ and $R_{FB}$ are typically a few kΩ, $R_S$ a few ohms, and $R_p$ ideally below 1 ohm. All of the circuit except for the room-temperature feedback electronics is at 300 mK.

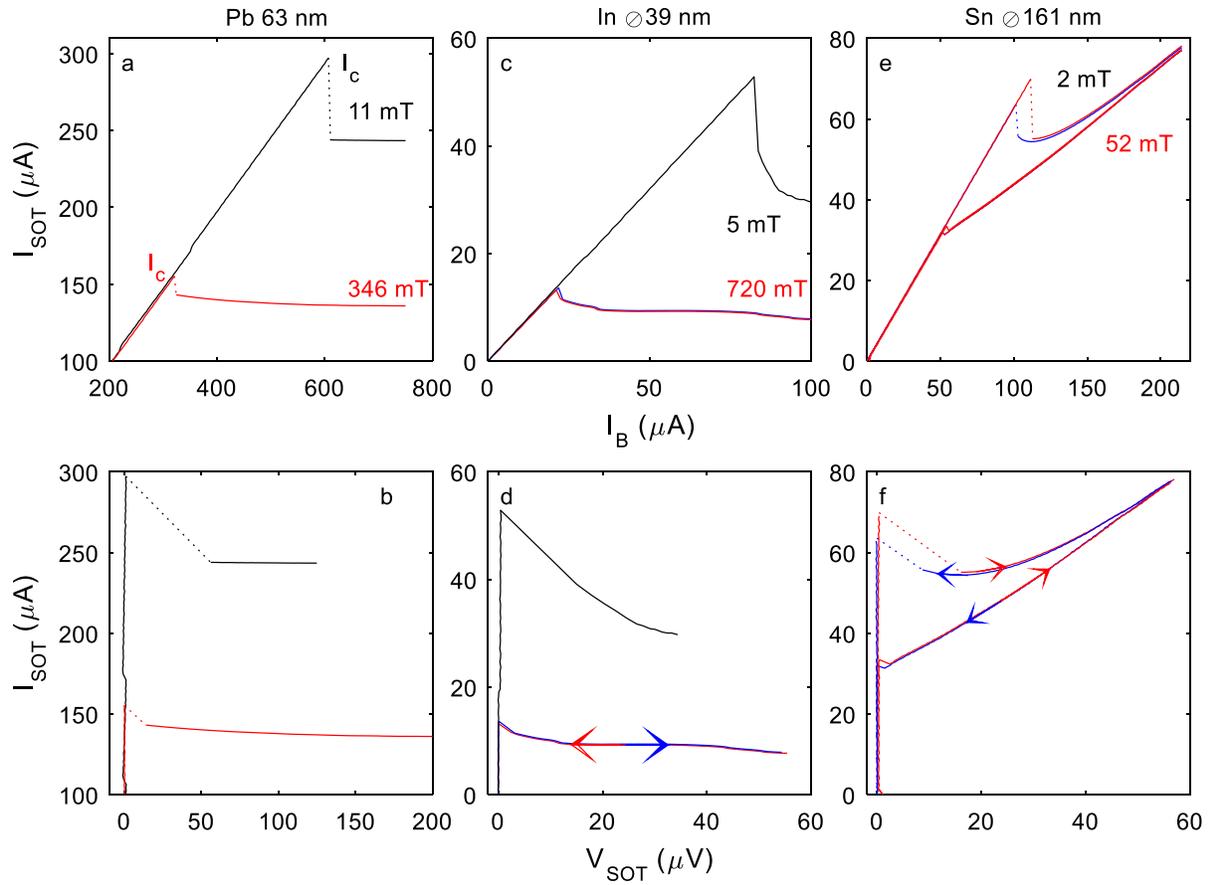

**Figure S2 - *I-V* characteristics of SOT devices made of different materials at 300 mK.** (a-f) $I$-$V$ characteristics of three different SOT devices at two different applied magnetic fields. (a) $I_{SOT}$-$I_B$ curve of a hysteric unshunted Pb SOT with diameter of 63 nm ($R_B = 2.0$ k$\Omega$, $R_s = 3.0\,\Omega$, $R_P = 3.1\,\Omega$). (b) The corresponding $I_{SOT}$-$V_{SOT}$ characteristics of the Pb device. (c) $I_{SOT}$-$I_B$ characteristics of an In SOT of 39 nm diameter with integrated shunt-on-tip ($R_B = 7.0$ k$\Omega$, $R_s = 3.0\,\Omega$, $R_P = 1.8\,\Omega$) and its corresponding $I_{SOT}$-$V_{SOT}$ characteristics (d). The red (blue) curve shows the measurement upon ramping $I_B$ up (down). (e) $I_{SOT}$-$I_B$ characteristics of a slightly hysteretic shunted Sn SOT with diameter of 162 nm ($R_B = 7.0$ k$\Omega$, $R_s = 3.0\,\Omega$, $R_P = 1.8\,\Omega$) with its corresponding $I_{SOT}$-$V_{SOT}$ characteristics (f). The inset shows a simplified diagram of the measurement circuit.

## Additional Fe$_3$O$_4$ nanocube images at zero field

Small particles such as Fe$_3$O$_4$ nanocubes can have defects that might influence the observations in this work. Here we show four other nanocubes that were imaged with the same 150 nm diameter In SOT as in the main text at the same temperature of about 300 mK. These images were also taken after ramping the field to 0.36 T. The out-of-plane field $B_z(x,y)$ images are shown in Fig. S3 a-c. For each of the images, the magnitude of the magnetic moment, $m_0$, as well as its polar, $\theta$, and azimuthal, $\phi$, angles were derived by performing a numerical fit of the data. Our model comprises a finite size SQUID loop with fitting parameters $m_0$, $\theta$, $\phi$ and the SOT-to-nanocube distance. The corresponding numerically attained best-fit stray field maps are presented in Figs. S3 d-f. The subtraction of the experimental and the corresponding best-fit image to show the quality of the fit is shown in Figs. S3 g-i. The magnitude and the polar angle are consistent with the finding shown in the main text. We have no evidence of other nanocubes behaving in a significantly different manner.

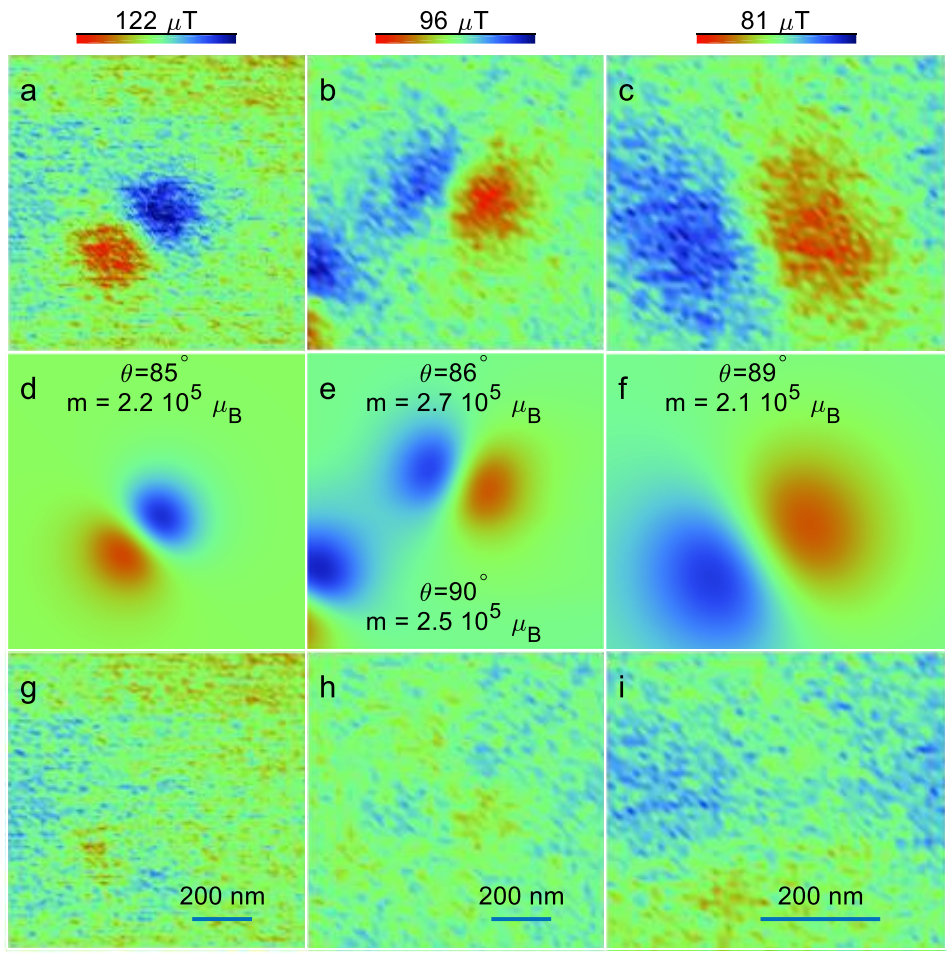

**Figure S3 - Additional images at zero field** – (a-c) Stray field $B_z(x,y)$ of different Fe$_3$O$_4$ nanocube at $\mu_0 H_z = 0$ T. (d-e) Corresponding best fit numerically simulated $B_z(x,y)$ with magnetic moment and polar angle found for each nanocube. The best tip-to-sample distance fit was 141 (d), 171 (e) and 165 nm (f) (g-h) Subtraction of the experimental image and the corresponding fit.